\documentclass[prl,twocolumn,tightenlines,footinbib]{revtex4-1}
\usepackage{bm,dcolumn,amsmath,graphicx,amsfonts,amssymb}

\usepackage{hyperref}
\usepackage{xcolor}
\definecolor{cite}{rgb}{0.,0.,0.9}   
\hypersetup{colorlinks,linkcolor={cite},citecolor={cite},urlcolor={cite}}

\newcommand{\bra}[1]{\ensuremath{\langle #1|}}	
\newcommand{\ket}[1]{\ensuremath{|#1\rangle}}	

\newcommand{\en}{\ensuremath{\varepsilon}}

\usepackage[normalem]{ulem}
\newcommand{\old}[1]{{\color{red}\sout{#1}}} %
\definecolor{newc}{rgb}{0.,0.6,0.4}

\newcommand{\note}[1]{{\color{orange}{${}^*$({#1})}}}

\renewcommand{\note}[1]{}
\renewcommand{\old}[1]{}

\renewcommand{\section}[1]{{\textit{\textbf{ #1}}---}}

\begin{document}

\title{Comment on ``New physics constraints from atomic parity violation in $^{133}$Cs''}

\author{B.\ M.\ Roberts}\email[]{b.roberts@uq.edu.au}
\author{J.\ S.\ M.\ Ginges}\email[]{j.ginges@uq.edu.au}
\affiliation{School of Mathematics and Physics, The University of Queensland, Brisbane QLD 4072, Australia}
\date{\today}

\begin{abstract}\noindent
In a recent Letter, B.\ K.\ Sahoo, B.\ P.\ Das, and H.\ Spiesberger, Phys.\ Rev.\ D {\bf103}, L111303 (2021)~\cite{Sahoo2021}, a calculation of the parity violating $6S-7S$ E1 amplitude in Cs is reported, claiming an uncertainty of just 0.3\%.
In this Comment, we point out that  key contributions have been omitted, and the theoretical uncertainty has been significantly underestimated.
In particular, the contribution of missed QED radiative corrections amounts to several times the claimed uncertainty.
\end{abstract}

\maketitle

\noindent
The $6S$--$7S$ atomic parity violation (APV) amplitude in Cs may be expressed as $\bra{\widetilde{7S}}D_z\ket{\widetilde{6S}}$, where $D_z$ is the 
$z$ component of the electric dipole (E1) operator,
and $\ket{\widetilde{6S}}$ and $\ket{\widetilde{7S}}$ are weak-interaction-perturbed atomic states; the source of this interaction is $Z$-boson exchange between the electrons and the nucleus. 
In the lowest-order single-particle picture,  it may be written
\begin{align}\label{eq:pnc}
{E}_{\rm PV} &= \sum_n\left[ \frac{\bra{7s}h_w\ket{n}\bra{n}d_z\ket{6s}}{\varepsilon_{7s}-\varepsilon_{n}}+\frac{\bra{7s}d_z\ket{n}\bra{n}h_w\ket{6s}}{\varepsilon_{6s}-\varepsilon_{n}}\right],
\end{align}
where 
$d_z$ is the single-particle E1 operator,
$h_w = -\frac{G_F}{2\sqrt{2}}Q_w \rho(r)\gamma_5$ is the parity-violating weak interaction operator, with $G_F$ the Fermi constant, $Q_w$ the nuclear weak charge, $\rho$ the nuclear density, and $\gamma_5$ the Dirac matrix, 
and $n$ runs over all $p_{1/2}$ states including the (occupied) core; see Ref.~\cite{GingesRev2004}.
The accuracy of the calculation is determined by account of many-body effects and smaller corrections
including higher-order relativistic effects.
Evaluation of $E_{\rm PV}$ in Cs with an accuracy matching or exceeding that of the measurement~\cite{Wieman1997} remains a formidable challenge.
There is a rich history connected to this spanning more than 20 years as the theoretical accuracy has reached the fraction-of-a-percent level; see, e.g., reviews~\cite{GingesRev2004,RobertsReview2015,AtomicReview2017} and Ref.~\cite{TohBeta2019}.
A major development over this time, following the realization of the significance of the Breit contribution~\cite{Derevianko2000,Derevianko2001,KozlovCs2001,DzubaCs2001},  was the recognition of the importance of quantum electrodynamics (QED) radiative corrections and the formulation of methods to account for them in precision calculations for heavy atoms~\cite{Sushkov2001,Johnson2001,GingesCs2002,Milstein2002,*Milstein2002a,*Milstein2003,Kuchiev2002a,*Kuchiev2002b,*KuchievQED03,
Sapirstein2003a,ShabaevPRL2005,FlambaumQED2005} (see also~\cite{Sapirstein2005,Shabaev2013,Ginges2016,GingesQED2015}).

We have identified a number of shortcomings in the theoretical evaluation of $E_{\rm PV}$ in the Letter~\cite{Sahoo2021}, some of which are detailed below.
Most notably, the treatment of QED radiative corrections omits important contributions to $E_{\rm PV}$, which amount to {\em several times} the theoretical uncertainty claimed in Ref.~\cite{Sahoo2021}.

\section{QED correction to $E_{\rm PV}$}
QED radiative corrections in the strong Coulomb field of the nucleus make a significant contribution to $E_{\rm PV}$, $\lesssim$\,1\%.
These have been calculated before~\cite{Johnson2001,GingesCs2002,Milstein2002,*Milstein2002a,*Milstein2003,
Kuchiev2002a,Kuchiev2002b,*KuchievQED03,
Sapirstein2003a,ShabaevPRL2005,Shabaev2005,FlambaumQED2005,RobertsQED2013} and are well established. 
It is said in the Letter~\cite{Sahoo2021} that one of the key improvements is the treatment of these QED corrections. 
However, details of the QED calculation are not presented in the Letter, and the reader is directed to the unpublished manuscript~\cite{Sahoo2020} for explanation~\footnote{Note that the reference to \cite{Sahoo2020} in the Letter~\cite{Sahoo2021} is incorrect, linking to an unrelated arXiv paper}.
There it is said that the self-energy QED correction to $E_{\rm PV}$ (and to other atomic properties) is accounted for by including the {\em radiative potential}~\cite{FlambaumQED2005,Ginges2016} into the Hamiltonian from the start~\footnote{Vacuum polarization is included using the standard Uehling potential, and a simplified form of the Wichmann-Kroll potential from Ref.~\cite{GingesCs2002,FlambaumQED2005}.},  which the authors claim to be a more rigorous approach compared to previous calculations.

\begin{figure}
\includegraphics[width=0.1\textwidth]{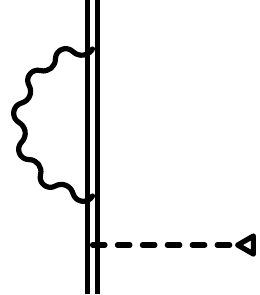}~~~~~
\includegraphics[width=0.1\textwidth]{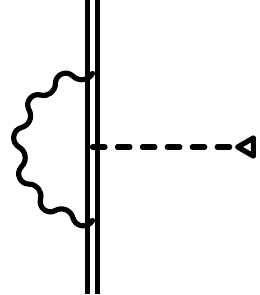}~~~~~
\includegraphics[width=0.1\textwidth]{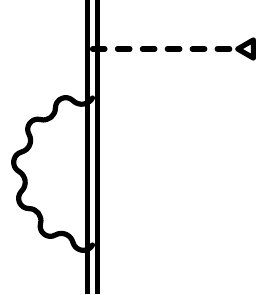}
\caption{Feynman diagrams for self-energy corrections to matrix elements. Dashed line with triangle represents the external field (e.g., E1, weak, hyperfine), wavy line the photon propagator, and double line the bound electron wavefunction and propagator. Middle diagram is vertex correction.}
\label{fig:qed}
\end{figure}

The radiative potential method~\cite{FlambaumQED2005} enables the accurate inclusion of self-energy corrections to the energies and wavefunctions of many-electron atoms. 
It may also be used to account for QED corrections to matrix elements of external fields whose operators act at radial distances much larger than the electron Compton wavelength, $r\gg e\hbar/(m_ec)$,  
e.g., the E1 field. 
However, this is not the case for operators that act at small distances, including the weak and hyperfine interactions. 
We illustrate this in Fig.~\ref{fig:qed}.
For the E1 interaction, the dominant contribution is given by the left and right diagrams, which may be accounted for by using the radiative potential method.
However, for the weak and hyperfine interactions, other contributions are important.
In particular, the middle vertex diagram -- where the external field is locked inside the photon loop -- simply cannot be accounted for using this method.
We refer the reader to the original~\cite{FlambaumQED2005} and subsequent~\cite{RobertsQED2013} works for details on the applicability of the radiative potential method.

The QED correction to the full Cs APV amplitude (involving both E1 and weak interactions) was determined in Refs.~\cite{FlambaumQED2005,ShabaevPRL2005}.
In Ref.~\cite{FlambaumQED2005}, the radiative potential method was used to calculate corrections to the E1 matrix elements and energy denominators in the sum \eqref{eq:pnc}, with QED corrections to weak matrix elements $\bra{s}h_w\ket{p_{1/2}}$ taken from previous works~\cite{Milstein2002,*Milstein2002a,*Milstein2003,Kuchiev2002a,Kuchiev2002b,*KuchievQED03,
Sapirstein2003a}.
In Ref.~\cite{ShabaevPRL2005}, Shabaev {\sl et al.}\ calculated the total correction by applying a rigorous QED formalism.
The results of Refs.~\cite{ShabaevPRL2005} and \cite{FlambaumQED2005} are in excellent agreement,  $-0.27(3)$\% and $-0.32(3)$\%, respectively~\footnote{The QED results of both Refs.~\cite{ShabaevPRL2005} and \cite{FlambaumQED2005} were misattributed in Table III of the Letter~\cite{Sahoo2021}; these papers were not cited in Ref.~\cite{Sahoo2021}.}.

It is unclear how the authors of~\cite{Sahoo2021} arrive at a QED correction of $-0.4\%$ for the weak matrix elements and $-0.3\%$ for $E_{\rm PV}$, in agreement with existing calculations~\cite{Milstein2002,*Milstein2002a,*Milstein2003,
Kuchiev2002a,Kuchiev2002b,*KuchievQED03,
Sapirstein2003a,FlambaumQED2005,ShabaevPRL2005,Shabaev2005,RobertsQED2013},
given the important short-range effects, including the vertex contribution, have been omitted.
In an attempt to reproduce the results of Ref.~\cite{Sahoo2021}, we calculate the radiative potential value for the QED correction to weak matrix elements, including vacuum polarization.
The result is $-2.1\%$, too large by a factor of five compared to the correct calculations, demonstrating the importance of the missed short-range effects.
This difference amounts to a change in $E_{\rm PV}$ that is nearly {\em six times} the atomic theory uncertainty claimed in Ref.~\cite{Sahoo2021}.

\section{Hyperfine constants}
In the Letter~\cite{Sahoo2021}, calculations of hyperfine constants are performed to test the accuracy of the wavefunctions in the nuclear region, crucial for assessing the accuracy of APV calculations  
(see Refs.~\cite{Ginges2017,Ginges2018,RobertsFr2020,RobertsHFA2021} for recent studies of the nuclear magnetization distribution for Cs).
By demonstrating excellent agreement with experiment, the authors conclude the accuracy of their wavefunctions is high, and so estimate a tremendously small uncertainty for the APV calculation. 
However, it appears that serious omissions have been made in the hyperfine calculations. 

As for $E_{\rm PV}$, the vertex and short-range contributions to QED corrections to hyperfine constants are important~\cite{Sapirstein2003b,Ginges2017} (see also~\cite{Blundell1997,Sunnergren1998,Artemyev2001,volotka08a,Volotka2012}).
Moreover, the magnetic loop vacuum polarization correction also gives a significant contribution~\cite{Sapirstein2003b,Ginges2017}.
In the Letter~\cite{Sahoo2021}, the radiative potential method is employed, with no account for these contributions.
Given this, it is unclear how the authors of~\cite{Sahoo2021,Sahoo2020} arrive at a correction of $-0.3\%$ to the hyperfine constants for $s$ states of Cs, in good agreement with existing 
calculations~\cite{Sapirstein2003b,Ginges2017}. 
To investigate this result, we again use the radiative potential method and find it gives
a correction of $-1.2\%$, three times too large compared to rigorous QED calculations~\cite{Sapirstein2003b,Ginges2017}, confirming the importance of the omitted effects.
This difference amounts to two times the uncertainty of the hyperfine calculations (0.4\%) claimed in the Letter~\cite{Sahoo2021}.



\section{Core contribution}
The contribution to $E_{\rm PV}$ coming from the (occupied) $n$\,=\,2--5 terms in Eq.~\eqref{eq:pnc} is called the ``core'' 
(or autoionization) 
contribution.
In the Letter~\cite{Sahoo2021}, it is said that 
the main difference in the $E_{\rm PV}$ result compared to the previous calculation of Dzuba {\em et al.}~\cite{OurCsPNC2012} stems from the opposite sign of the core contribution.
The difference in core contribution between Refs.~\cite{Sahoo2021} and \cite{OurCsPNC2012} is larger than the theoretical uncertainty claimed in the Letter~\cite{Sahoo2021}
and should be investigated thoroughly.

In Ref.~\cite{OurCsPNC2012}, Dzuba {\em et al.}\ showed that many-body effects (core polarization and correlations) have a significant impact on the core contribution, changing its sign compared to the lowest-order Hartree-Fock value; see also Ref.~\cite{RobertsReview2015}.
The authors of Ref.~\cite{Sahoo2021} claim their result confirms the core calculation of Ref.~\cite{Porsev2009,*Porsev2010} and agrees with the result of Ref.~\cite{Blundell1990,*Blundell1992}. 
However, in both of those works, the core contribution was evaluated in the lowest-order approximation.

Here, we re-examine the core contribution in detail in an attempt to elucidate the source of this discrepancy.
We include core polarization using the time-dependent Hartree-Fock (TDHF) method~\cite{DzubaHFS1984}, in which the single-particle operators are modified:
$d_z \to {\tilde d_z}  = d_z+\delta V_{d}$, and
${h_w} \to {\tilde h_w}={h}_w+\delta V_{w}$.
The $\delta V$ corrections are found by solving the set of TDHF equations for all electrons in the core~\cite{DzubaHFS1984}. 
We obtain the corrections to lowest-order in the Coulomb interaction by solving the set of equations once, 
and to all-orders by iterating the equations until self-consistency is reached~\cite{DzubaHFS1984} (equivalent to the random-phase approximation with exchange, RPA~\cite{Johnson1980}).
The equations for $\delta V_d$ are solved at the frequency of the $6S$--$7S$ transition
(see \cite{RobertsDCP2013} for a numerical study). 
%
We account for correlation corrections 
using the second-order~\cite{Dzuba1987jpbRPA} and all-orders~\cite{DzubaCPM1988pla,*DzubaCPM1989plaEn,*DzubaCPM1989plaE1} correlation potential methods (see also \cite{GingesCs2002}).

The core contribution arises as the sum of two terms, due to the weak-perturbation of $6s$ and $7s$ states, respectively.
These have similar magnitude though opposite sign, and strongly cancel,
meaning numerical error may be significant.
We test the numerical accuracy in a number of ways.
Firstly, we vary the number of radial grid points used for solving the differential equations, and vary the number of basis states used in any expansions.
We find numerical errors stemming from grid/basis choices can easily be made insignificant.
More importantly, we have three physically equivalent, but numerically distinct, ways to compute $E_{\rm PV}$:
\begin{align}
&\sum_n \left[\frac{\bra{7s}\tilde h_w\ket{n}\bra{n}{\tilde d_z}\ket{6s}}{\varepsilon_{7s}-\varepsilon_{n}}
+\frac{\bra{7s}{\tilde d_z}\ket{n}\bra{n}\tilde h_w\ket{6s}}{\varepsilon_{6s}-\varepsilon_{n}} \right] \label{eq:sos}\\
&= \bra{\delta\psi_{7s}}{\tilde d_z}\ket{6s} + \bra{7s}{\tilde d_z}\ket{\delta\psi_{6s}} \label{eq:se-h} \\
&= \bra{7s} \tilde h_w\ket{\Delta\psi_{6s}} + \bra{\Delta\psi_{7s}}\tilde h_w\ket{6s} , \label{eq:se-d}
\end{align}
where $\delta \psi$ and $\Delta \psi$ are corrections to the valence wavefunctions ($\psi$) due to the 
time-independent weak interaction, 
and the 
time-dependent E1 interaction, 
respectively.
These are called the sum-over-states (\ref{eq:sos}), weak-mixed-states (\ref{eq:se-h}), and E1-mixed-states (\ref{eq:se-d}) methods~\footnote{These formulas exclude the double-core-polarization effect, which is very small for Cs, and has been studied in detail in Ref.~\cite{RobertsDCP2013}. The sign change of the core cannot be explained by the double-core-polarization correction, which even if entirely assigned to the core contribution is a factor of two too small to account for the difference~\cite{RobertsDCP2013}.}.

In the sum-over-states method, a B-spline basis (e.g., \cite{Johnson1988,Beloy2008}) is used to sum over the set of intermediate states. 
In contrast, the mixed-states approach does not require a basis at all; the $\delta$ and $\Delta$ corrections are found by solving the differential equations~\cite{Dalgarno1955}: 
\begin{align}
(h - \en )\delta\psi &= -{\tilde h_w}\psi
\\
(h - \en  - \nu)\Delta\psi &= -{\tilde d_z} \psi,
\end{align}
where $h$ is the single-particle atomic Hamiltonian, and $\nu$ is the $6S$--$7S$ transition frequency.
In the mixed-states approach, the core contribution is found by projecting the corrections $\delta\psi$ and $\Delta\psi$ onto the core states, while in the sum-over-states method it is found by restricting the sum to include only core states.
Note that the numerics involved in solving each of the above equations is significantly different, and the coincidence of results is indicative of high numerical accuracy.
Even with moderate choice for the radial grid, we find the results of the two mixed-states methods agree to parts in $10^8$, and the mixed-states and sum-over-states methods agree to parts in $10^7$, demonstrating excellent numerical precision and completeness of the basis.

Our calculations of the core term are summarized in Table~\ref{tab:core}.
The sign change in the core contribution is mostly due to polarization of the core by the external E1 field.
This is sensitive to the frequency of the E1 field.
While correlations beyond core polarization are important, they affect both terms in roughly the same manner; the core term and its sign are robust to the treatment of correlations.
We also performed calculations for the $7S$-$6D_{3/2}$ $E_{\rm PV}$ for $^{223}$Ra$^+$ to test against previous calculations;
at the RPA level, we find the core contribution to be 6.81 [in units $-10^{-11}i(-Q_w/N)\,|e|a_B$], in excellent agreement with the result 6.83 of Ref.~\cite{Pal2009}
(see also~\cite{DzubaPNCsd2001,Wansbeek2008}).
%
It is unclear why the sign of the result of Ref.~\cite{Sahoo2021} remains the same as the Hartree-Fock value, however, we note that it may not be straight forward to compare individual contributions across different methods as discussed in Refs.~\cite{Wieman2019,AtomicReview2017}.

\section{Conclusion}
For the above reasons, we are not convinced the result presented in the Letter~\cite{Sahoo2021} is an improved value for the Cs $E_{\rm PV}$. 
We conclude that the most reliable and accurate values that have been obtained to date are:
$E_{\rm PV}=0.898(5)$~\cite{GingesCs2002,FlambaumQED2005} and $E_{\rm PV}=0.8977(40)$~\cite{Porsev2009,OurCsPNC2012}, in units $-10^{-11}i(-Q_w/N)\,|e|a_B$, 
which agree precisely and were obtained using different approaches.
These results are also in excellent agreement with previous calculations \cite{DzubaCPM1989plaPNC,Blundell1990,*Blundell1992,KozlovCs2001,Shabaev2005}, though in disagreement with the result of the Letter~\cite{Sahoo2021}.

\begin{table}
\caption{Core contribution to $^{133}$Cs 6S-7S $E_{\rm PV}$ in different approximations, in units  $-10^{-11}i(-Q_w/N)\,|e|a_B$, where $N=78$ is the number of neutrons.$^a$
Here, HF denotes relativistic Hartree-Fock,
$\delta V^{(1)}$ and $\delta V^{(\infty)}$ denote lowest-order and all-orders core-polarization, respectively, with subscripts $w$ and $d$ indicating polarization by the weak or E1 fields, 
$\Sigma^{(2)}$ and $\Sigma^{(\infty)}$ denote second- and all-orders correlations, respectively, and $\lambda$ indicates correlations have been re-scaled to reproduce the lowest experimental binding energies.
\label{tab:core}}
\begin{ruledtabular}
\begin{tabular}{l D{.}{.}{2.5}D{.}{.}{2.5}D{.}{.}{2.5}}
\multicolumn{1}{c}{Method}& 
\multicolumn{1}{c}{$\bra{\delta\psi_{7s}}{\tilde d_z}\ket{6s}$} &
\multicolumn{1}{c}{$\bra{7s}{\tilde d_z}\ket{\delta\psi_{6s}}$} &
\multicolumn{1}{c}{Sum} \\
\hline
HF      & -0.02645 & 0.02472 & -0.00174 \\
HF+$\delta V_{w}^{(1)}$     & -0.03747 & 0.03539 & -0.00208 \\
HF+$\delta V_{w}^{(\infty)}$    & -0.04319 & 0.04119 & -0.00201 \\
\hline
\multicolumn{4}{c}{E1 TDHF equations solved at HF frequency:}\\
HF+$\delta V_{w}^{(\infty)}$+$\delta V_d^{(1)}$ & -0.05506 & 0.05442  & -0.00063 \\
HF+$\delta V_{w}^{(\infty)}$+$\delta V_d^{(\infty)}$\tablenotemark[2] & -0.05822 & 0.05992 & 0.00170  \\
\hline
\multicolumn{4}{c}{E1 TDHF equations solved at experimental frequency:}\\
HF+$\delta V_{w}^{(\infty)}$+$\delta V_d^{(1)}$ & -0.05468 & 0.05466  & -0.00002\\
HF+$\delta V_{w}^{(\infty)}$+$\delta V_d^{(\infty)}$\tablenotemark[2] & -0.05784 & 0.06043 & 0.00259  \\
\hline
\multicolumn{4}{c}{Including correlation corrections (and $\delta V_{w}^{(\infty)}+\delta V_d^{(\infty)}$):}\\
$\Sigma^{(2)}$ & -0.06739 & 0.06924 & 0.00184\\  
$\lambda\Sigma^{(2)}$ & -0.06547 & 0.06732 & 0.00184\\
$\Sigma^{(\infty)}$ & -0.06514 & 0.06695 & 0.00181\\ 
$\lambda\Sigma^{(\infty)}$ & -0.06516 & 0.06696 & 0.00181\\
\hline
\multicolumn{4}{c}{Other calculations:}\\
HF~\cite{Blundell1990,*Blundell1992}  & & &-0.002\\
HF~\cite{Porsev2009,*Porsev2010} & & &-0.002\\
$\Sigma^{(\infty)}$+RPA~\cite{OurCsPNC2012}  & & &0.00182\\
\hline
\multicolumn{4}{c}{Values from the Letter~\cite{Sahoo2021}:}\\
HF~\cite{Sahoo2021}   && &-0.0017\\
RCCSD~\cite{Sahoo2021} & & &-0.0019\\
RCCSDT~\cite{Sahoo2021}  & & &-0.0018\\
\end{tabular}
\end{ruledtabular}
\tablenotetext[1]{To avoid possible ambiguity in the sign, we note that the total amplitude is positive in these units;
at the HF level it is $0.7395$.}
\tablenotetext[2]{HF+$\delta V_{w}^{(\infty)}$+$\delta V_d^{(\infty)}$ is commonly called RPA level.}
\end{table}


{\em Acknowledgments---}
We thank V.~A.~Dzuba and V.~V.~Flambaum for useful discussions.
This work was supported by the Australian Government through ARC DECRA Fellowship DE210101026 and ARC Future Fellowship FT170100452.

\bibliography{library}

\end{document}